\documentclass[conference]{IEEEtran}
\IEEEoverridecommandlockouts
\usepackage[noend]{algorithm}         
\usepackage[noend]{algorithmic}     
\usepackage{multirow}
\usepackage{enumerate}
\usepackage{xcolor}
\usepackage{bm}
\usepackage{booktabs}
\usepackage{subfigure}
\usepackage{hyperref}
\usepackage{color}
\usepackage{cite}
\usepackage{amsmath,amssymb,amsfonts}
\usepackage{algorithmic}
\usepackage{graphicx}
\usepackage{textcomp}

\def\BibTeX{{\rm B\kern-.05em{\sc i\kern-.025em b}\kern-.08em
    T\kern-.1667em\lower.7ex\hbox{E}\kern-.125emX}}
\begin{document}

\title{Population Density-based Hospital Recommendation \\ with Mobile LBS Big Data \\
\thanks{This work is supported in part by National Natural Science Foundation of
China (NSFC) (Grant No. 61673118) and in part by Shanghai Pujiang Program
(Grant No. 16PJD009).}
}

\author{
\IEEEauthorblockN{
Hanqing Chao\IEEEauthorrefmark{1},
Yuan Cao\IEEEauthorrefmark{1},
Junping Zhang\IEEEauthorrefmark{1},
Fen Xia\IEEEauthorrefmark{2},
Ye Zhou\IEEEauthorrefmark{1}, and
Hongming Shan\IEEEauthorrefmark{1}}
\IEEEauthorblockA{\IEEEauthorrefmark{1}Fudan University, 220 Handan Road, Yangpu District, Shanghai 200433 China 
                    \\ \{hqchao16, caoy16, jpzhang, yezhou14, hmshan\}@fudan.edu.cn}
\IEEEauthorblockA{\IEEEauthorrefmark{2}Beijing Wisdom Uranium Technology Co., Ltd.  xiafen@ebrain.com}
}

\maketitle

\begin{abstract}
The difficulty of getting medical treatment is one of major livelihood issues in China. Since patients lack prior knowledge about the spatial distribution and the capacity of hospitals, some hospitals have abnormally high or sporadic population densities. This paper presents a new model for estimating the spatiotemporal population density in each hospital based on location-based service (LBS) big data, which would be beneficial to guiding and dispersing outpatients. To improve the estimation accuracy, several approaches are proposed to denoise the LBS data and classify people by detecting their various behaviors. In addition, a long short-term memory (LSTM) based deep learning is presented to predict the trend of population density. By using Baidu large-scale LBS logs database, we apply the proposed model to 113 hospitals in Beijing, P. R. China, and constructed an online hospital recommendation system which can provide users with a hospital rank list basing the real-time population density information and the hospitals' basic information such as hospitals' levels and their distances. We also mine several interesting patterns from these LBS logs by using our proposed system. 
\end{abstract}


\section{Introduction}\label{sec:intro}

\IEEEPARstart{A}{ccording} to the statistics of National Health and Family Planning Commission (\href{http://www.nhfpc.gov.cn/mohwsbwstjxxzx/s7967/201702/0a644a51bfc347ccab43fb1766aa5089.shtml}{https://goo.gl/i2Kh6p}), there are a total of 991,632 medical institutions in China until Nov. 2016 where 4735 new institutions are increased from Nov. 2015. Nevertheless, the difficulty of getting medical care remains as one of China's major livelihood issues. A survey on Peking University First Hospital~\cite{yu2008survey}, which is one of the most famous hospitals in Beijing, indicates that more than 45\% outpatients have to wait for over two hours after the registration, whereas 85\% of them have less than 10 minutes for the doctor's inquiry. This phenomenon is actually common in China's 776 Top-Class hospitals\footnote{Chinese government classifies all hospitals in China into three levels and each level is divided into three subclasses, i.e. 9 classes in total.}. The reason is that many hospitals lack publicity and are not familiar to most patients. Having no way of knowing whether there is a good enough and less crowded hospital nearby, people have no choice but go to those congested famous hospitals regardless of the severity of the disease. In fact, most outpatients with a mild disease may expect a quick treatment yet have a low requirement for the hospital's treatment ability. It is imperative to find a simple and convenient way to access the crowd status and basic information of the neighboring hospitals. There are several standard ways for crowd counting, such as approaches based on video or beacon. However, these methods rely on surveillance data or wireless network data, and any company or non-governmental organization can hardly gather these data of all hospitals even in one city, not to mention in any larger range.

Fortunately, location-based service (LBS) big data offers a potential solution to this dilemma. LBS data have two unique properties: \textbf{a)} LBS data naturally belongs to the service providers. Using it for population density estimation does not involve asking hospitals for any help. \textbf{b)} LBS data is sufficient for population density estimation with its copiousness and vast area coverage. As smartphone is popular, several main LBS providers in China have hundreds of millions of users and preserve billions LBS request logs every day. For instance, Baidu Map has over 300 million active users and gets 23 billion LBS request logs in China each day on average. Taking advantage of these two properties, we present a novel (nearly) real-time model for counting and predicting the crowd of all hospitals in the city based on LBS big data. What's more, the distribution of people's residence time can also be figured out with the LBS data. Relying on the location logs of Baidu LBS and Baidu Points of Interests (POI) data, we have designed an APP\footnote{In our application, we use logs located in Beijing, P.R. China, which is in the order of $10^8$ per day. The exact amount of these logs is sensitive information for Baidu co. which cannot be revealed.} to provide users with \textbf{a)} hospitals' real-time and predicted outpatient density, \textbf{b)} official level, \textbf{c)} distance from the user to each hospital and the corresponding path planning. To the best of our knowledge, it is the first hospital recommendation system based on the population density analysis with LBS big data. Note that Baidu LBS data has incorporated the information of GPS, WiFi, and Cellular network, so it is a relatively high-quality data source. Since the LBS data we use is location logs of mobile APPs, we will use \emph{data} and \emph{logs} interchangeably to denote the LBS data.

Our contributions can be summarized into three innovations conquering three main challenges of this work. \emph{The first innovation} is the model for detecting the types of people around hospitals to address highly noisy LBS data. Although our data source is relatively high-quality, it is still temporally unstable and spatially inaccurate for a variety of reasons. Passersby, inpatients, and hospital staff, as well as people working around hospitals can all cause noise in our outpatient density estimation task. By identifying the different movement behaviors, the population analysis model clusters the users into three classes quickly, i.e., \textbf{a)} people who pass by, \textbf{b)} who are outpatients, and \textbf{c)} who are inpatients or staff. Since the goal is to provide references to outpatients with mild disease, only the outpatients should be counted. This work will be specified in Section~\ref{sec:PDA}. \emph{The second innovation} is a neural network for predicting the population density trend to improve the practicality of our application. Accurate estimations of how long patients need to wait for treatments are what patients actually concern. Based on the history statistics of population density obtained by the analysis of the LBS data, a dual network is designed for representing both high-frequency and low-frequency trend and generating accurate predictions. We will introduce it in Section~\ref{sec:PDP} \emph{The third innovation} is a highly active parallel construction on Hadoop. Providing real-time population density analysis and prediction requires the system to process dozens of Gigabyte information per hour. To address this issue, we fully utilize the features of Hadoop by separating the whole task into two MapReduce processes. Consequently, our application can process 20GB original data within 10 minutes by using 2000 slave nodes. Details of the construction are in Section~\ref{sec:parall}

\begin{figure}[htbp]
\centering
\includegraphics[width=0.8\linewidth, clip=true, trim=0 0 0 0]{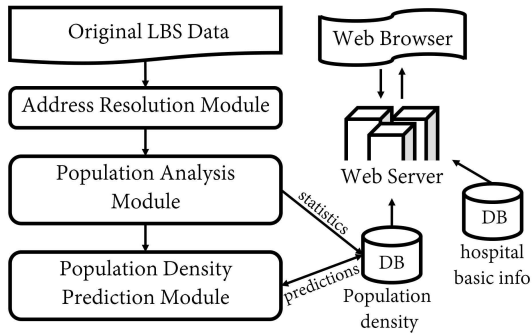}
\caption{The flowchart of our hospital recommendation system. Three major modules calculate the real-time population density of each hospital and predict its trend. The web server merges the population density information and hospital basic information for users.}
\label{fig:whole_structure}
\end{figure}

Figure \ref{fig:whole_structure} schematically shows how the innovations integrate into an application. There are three major modules, two databases, and a web server. \emph{Address resolution module} (AR) resolves the address of each piece of LBS log according to its longitude and latitude. \emph{Population analysis module} (PA) analyzes population density after removing the noisy data in the LBS logs. These two modules calculate statistics of population density. \emph{Population density prediction module} (PDP) estimates population density trend basing the history statistics. With these three modules cooperating with each other, the real-time population density data and its trend in the future several hours can be obtained. So, users can get the combination of population density information and hospitals' basic information from the web server .

\section{Related Work}\label{sec:RW}

In the industries of China, there have been three crowd-counting related trials based on LBS big data. One of them is Baidu's big data project on China's ghost cities~\cite{chi2015ghost}. The project discovered 50 possible ghost cities by analyzing the LBS data in a long-term scale. Since the strategy of address resolution in this work is exquisite, we referenced it in our work. But unlike this project~\cite{chi2015ghost}, we investigate the LBS data in a short-term scale and can output the results in nearly real time. Another two trials are heat maps of Baidu Map and Wechat, each of which encodes the population density by different colors. Both of these two works studied population density based on the holistic map, which restricts them to mining unique human behavior features in a certainly bordered area or a certain kind of public facility. By perusing these behavior features, we developed several optimization strategies to improve the reliability of results in our work. Functionally, our application presents the density level and density-time curves of each hospital, while Baidu Map only provides current heat map without density-time curves and Wechat City Heat Map gives a density-time curve of the selected point on the map instead of a geographical region.

Other related work on the application of big data to public healthcare was primarily focused on mining social media data for bio-surveillance systems~\cite{dredze2012social,velasco2014social,denecke2012making}, Electronic Health Record (EHR)~\cite{zhang2015clinical,jensen2012mining}, and locating health resources~\cite{pacheco2008heuristic,kim2012heuristics}. It is worth pointing out that most of the previous work aimed at healthcare facilities, whereas we focus on complementary beneficiaries. Our work is patients-oriented.

Moreover, the majority of the researches on Point-of-Interest (POI) recommendation and LBS or Location-based Social Networks (LBSNs) pay attention to four aspects, i.e., temporal patterns~\cite{yao2016poi,yin2015joint}, geographical influence~\cite{lian2014geomf,yin2015joint}, social correlations~\cite{hu2014social} and textual~\cite{vu2016geosocialbound} or visual~\cite{lim2015recommending,wang2017your} content indications. Instead of improving the recommendation approach, our research focuses on the estimation and prediction of population density based on massive LBS data which is a totally different database with LBSNs data. And we use a simple yet effective recommendation strategy to show the distinctive value of this work.

\section{Population Density Analysis}\label{sec:PDA}
In this section, we will show how to recognize different classes of people to gain denoised LBS data and analysis population density on that. Two modules are introduced, the \emph{address resolution module} (AR) and \emph{population analysis module} (PA). The PA module takes most of the analysis work, yet it is time-consuming. Hence, we present efficient AR module to pretreat the origin data and screen out enormous irrelevant data.

\subsection{Address Resolution (AR)}\label{subsec:AR}
Hospitals only occupy a small proportion of the whole city land. If we simply assume the LBS data located uniformly in the city, the natural inference is that majority of original data can be filtered out after its location is decoded. Thus, several strategies are proposed for address resolution as follows. First of all, we map the continuous longitude and latitude to a discrete set of squares of length $a$~\cite{chi2015ghost}. In other words, the map is cut into grids, each of which has a semantic information indicating {\bf a)} whether this grid belongs to a hospital and {\bf b)} if it does, which hospital it belongs to. After a compromise between precision and efficiency, the grid length $a$ is set to be 2 meters. Next, each grid is allocated a unique ID calculated by its center latitude and longitude. For a fast query, the map between grids' IDs and their semantic information is stored. In this manner, we can obtain the semantic location information of a specific log just by calculating the ID of its corresponding grid. After filtering out data irrelevant to hospitals, the number of logs is significantly reduced from the order of $10^8$ per day to the order of $10^6$ per day.

\subsection{Population Analysis (PA)}

To get accurate population, we propose a population classification strategy. In a typical public service area, different types of persons have remarkably diverse behaviors. By detecting and identifying these behaviors, we can accurately screen outpatients and denoise the original data. But before analyze behaviors, another work needs to be done. Although temporally unsteadiness and spatially inaccuracy are basic properties of LBS data and we cannot improve that, we introduce an auxiliary variable $c$ to describe the uncertainty of each piece of location log.

For each request log, there are observed latitude $lat_{o}$ and longitude $lng_{o}$ for describing the observation position and $r$ for observation accuracy. The formal definition of accuracy $r$ is that the point is guaranteed to be within the circle centered on $(lat_{o}, lng_{o})$ and with radius $r$. Therefore, let $\bm{x}$ be a random variable representing the true position of this log. We assume $\bm{x}$ follows an isotropic normal distribution $\mathcal{N}$ with the 2-dimensional mean vector $\bm{\mu}=(lat_{o}, lng_{o})$ and the $2\times2$ covariance matrix $\bm{\Sigma}=\sigma\bm{I}$, where parameter $\sigma=r/3$~\footnote{An isotropic bivariate normal distribution, $\bm{x}\sim\mathcal{N}(\bm{\mu}, \sigma\bm{I})$, has such property: $\mathbf{Pr}(-3\sigma \le \bm{x} \ge 3\sigma)>0.99$. So setting $\sigma$ as $r/3$ properly indicates the definition of $r$.} and $\bm{I}$ is a $2\times2$ identity matrix. Then, we describe the certainty that a request log is truly in the hospital by a confidence $c$:
\begin{equation}
c=\iint\limits_{\bm{R}}f(\bm{x})d\bm{x}\approx
\frac{|\bm{X'}\cap \bm{R}|}{|\bm{X'}|}
\end{equation}
where $\bm{R}$ is the region of the hospital this log belongs to and $f(\bm{x})$ is the probability density function of $\mathcal{N}(\bm{\mu}, \bm{\Sigma})$. To calculate $c$ efficiently, we sample $k$ points with the normal distribution $\mathcal{N}$ forming a point set $\bm{X'}$ and use the percentage of positive samples located in the hospital, to approximate $c$. According to the definition, $c\in[0,1]$ and $c=1$ indicates this log is surely located in this hospital.

\begin{figure}[htbp]
\centering
\includegraphics[width=1\linewidth, clip=true, trim=100 70 80 70]{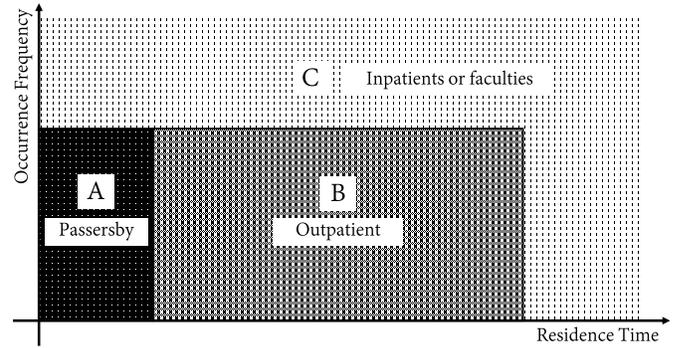}
\caption{Part A represents passersby who momentarily appeared. Part B represents outpatients. Part C characterizes people who stay for a very long time or appear regularly which indicates that they might be inpatients or staff}
\label{fig:fig1}
\end{figure}

With LBS logs with confidence $c$, we can apply our population classification. Based on our observation, the differences among behaviors mainly reflect in residence time and occurrence frequency, as shown in Figure~\ref{fig:fig1}. Hence, we screen out people who stay for a rational time and do not present frequently with some empirical thresholds and rules. About the residence time, on one hand, a patient couldn't finish the whole diagnostic procedure within 20 minutes, even eliminating all the queuing issues. On the other hand, a 20-minute walk can reach far more than 300 to 500 meters, which can be the maximal length of a hospital in a city. Based on this property, we regard people staying less than 20 minutes as passersby on the nearby road and filter these data out. Further, outpatients will not stay at a hospital for an extremely long time, for example, over 15 hours. Since outpatient doctors are on duty from 8 a.m. to 6 p.m., a total of 10 hours. With additional 5 hours for containing patients starting queuing before the work time and still in treatment like intravenous infusion after closed, 15 hours is a very loose upper bound for outpatients' duration. Those citizens staying for a longer time might be an inpatient. About the occurrence frequency, staff of a hospital or residents around can be depicted more clearly in this perspective. In a long term, an outpatient would unlikely go to the hospital every day, but staff or residents are quite the opposite. Precisely, people who are observed more than 3 times in consecutive 7 days will be considered as staff or residents. Overall, if one person's residence time $t$ meets $20 min\le t\le15 h$ and, meanwhile, he wasn't observed frequently, then he will be regarded as an outpatient. 

In particular, two structures, \emph{counting list} and \emph{blacklist}, are designed to analyze and count people. Each hospital has its corresponding \emph{counting list} and \emph{blacklist}. \emph{Counting list} is used for counting people and wipe out those who stay too short. \emph{Blacklist} will record those who stay too long or present frequently. We now explain each of these two structures.

\textbf{Counting list.} It stores the information of each person in this hospital at present. It is a hash list with each \emph{item} representing a person distinguished by UID. Each \emph{item} includes $3$ dimensions:
\begin{enumerate}[a)]
\item \emph{res\_time} recording the one's residence time.
\item \emph{confidence} $\hat{c}$ of this person.
\item \emph{is\_patient} indicating whether this person has stayed for enough time, initialized as $False$.
\end{enumerate}
Let $n^c$ be the number of \emph{item}s of one \emph{counting list}. Let $i$ be the index of each \emph{item}, $i\in\{1,2,..,n^c\}$. The number of people $N$ in this hospital at any time is estimated by formula: $N=\sum_{k\in T}\hat{c}_k$ where $T$ is the set of all $i$ meet $is\_patient_i=$ TRUE in this \emph{counting list}.

The residence time analysis strategy involves two variations, $is\_patient$ and $confidence\ c$:
\begin{align}
	&is\_patient_i=\mbox{TRUE}, &
		\begin{split}
		&\mbox{if }res\_time_i>20min
		\end{split} \\
	&\hat{c}_i:
	\begin{cases} 
	=(\hat{c}+c) \mod 1,\\
	=\hat{c}_i/2,  \\
	\mbox{delete } item_i,
	\end{cases} &
		\begin{split}
		&i\mbox{'s new data received}\\
		&\mbox{every 15 min}\\
		&\hat{c}_i<2^{-16}
		\end{split}
\end{align}
where. Since when patients leave the hospital, we cannot receive any special signal, the exponential decay strategy of $\hat{c}$ can improve the accuracy of population counting. And the deleting condition simply indicates when we haven't received a person's LBS log for 4 hours, we believe this person has left the hospital. 

\textbf{Blacklist.} People who are classified as inpatients or staff or residents around are kept in \emph{blacklist}. If a LBS log belongs to the person in the \emph{blacklist}, it will be ignored. \emph{Blacklist} also has a deleting rule. If a person didn't show up in this hospital for consecutive 10 days, he will be removed.

\section{Population Density Prediction}\label{sec:PDP}

In Chinese hospitals, the schedule of each doctor is almost fixed every week. Therefore, it is natural to assume that the hospital population density varies periodically according to one-week cycle. Considering such a variation is time-series, we expect to predict the population density trend from the historical data by utilizing neural network based on long short-term memory deep learning (LSTM)~\cite{hochreiter1997long}. Note that because the population densities of different hospitals have their respective changing trends, we thus train the network for each hospital separately.


Generally, we want the prediction model to learn the pattern between the input historical data and the population density of next hours. Since the hospital population density varies periodically according to one-week cycle, every week's density-time curve looks similar to each other. Hence, we slice one week into several equal-length periods. Let $n^p$ be the total period number in a week. Each period is identified by a vector $\bm{p}=(w,t)$, where $w$ is the week index, $t\in{1,2,..,n^p}$ is the period index in the week. Given period $\bm{p}_{w,t}$ the population density data set in this is denoted by $\bm{D}_{\bm{p}}$ or $\bm{D}_{w,t}$. The periodicity mentioned above can be described as given $w$ and $t$, for $\forall w_1,w_2 \in \varepsilon$, $\bm{D}_{w_1,t}$ is similar to $\bm{D}_{w_2,t}$, where $\varepsilon$ is the neighborhood of $w$. Therefore, we expect the network can learn the function:
\begin{align}
\label{func:lstm}
\begin{cases} 
F(\bm{D}_{w-1,t+1},\bm{D}_{w,t})=\bm{D}_{w,t+1},\\
F(\bm{D}_{w,1},\bm{D}_{w,t})=\bm{D}_{w+1,1},
\end{cases} &
  \begin{split}
  t<n^p\\
  t=n^p
  \end{split}
\end{align}

According to the definition above, we can use a slid window for slicing periods from one-week data. The length of each period depends on the window length $l$ (hours). The distance between each pair of successive periods is determined by the step size $s$ (hours). How to select $l$ and $s$ is tricky. Because except referring to the historical data $\bm{D}_{w-1,t+1}$, the network needs to learn the pattern $f_{t}$ between two successive periods, $\bm{D}_{w,t}$ and $\bm{D}_{w,t+1}$. We naively present two hypotheses. \textbf{a)} For $\forall t_1,t_2 \in \{1,2,..,n^p\}$ and $t_1 \ne t_2$, $f_{t_1} \ne f_{t_2}$. \textbf{b)} If the cardinality of $D$ is smaller, the network tends to learn the more high-frequency pattern and have a greater possibility of overfitting. Conversely, the network tend to learn the more low-frequency pattern and have a greater possibility of underfitting. We now detail the effects of $l$ and $s$.

\begin{figure}[htbp]
\centering
\includegraphics[width=1\linewidth, clip=true, trim=240 165 285 190]{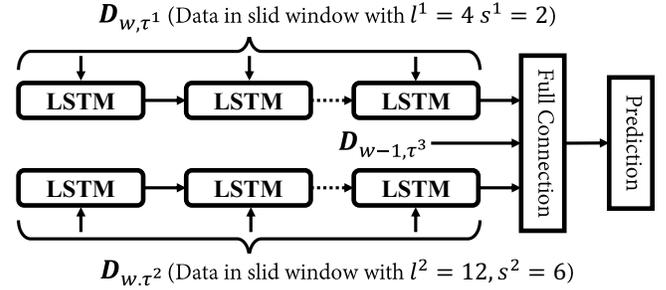}
\caption{The architecture of our dual network.}
\label{fig:lstm-m}
\end{figure}

\textbf{Step size $s$.} As one week have 7 days and one day have 24 hours, $n^p=24*7/s$. Hence, A smaller $s$ will generate more periods. According to hypothesis \textbf{a)}, there will be more different $f$ for the network to learn which might cause underfitting. However, a large $s$ brings a long time span between two successive predictions which can decrease the practicality of the application. what's more, to cover entire time range, $s$ should not be larger than $l$, i.e. $s \le l$. 

\textbf{Window length $l$.} Since statistic data set for every hour has the same cardinality, the cardinality of $D$ is directly determined by $l$. Along with hypothesis \textbf{b)}, the impact of $l$ is clear.

Instead of struggling between large or short window length and step size, we introduce a dual network to seek a balance. The structure is shown in Figure~\ref{fig:lstm-m}. Two period-slicing strategies are chosen for two independent LSTM \cite{hochreiter1997long} nets. One strategy chooses $l^1=4, s^1=2$. The other chooses $l^2=12, s^2=6$. Meanwhile, Formula~\ref{func:lstm} is improved. As there are two period-slicing strategies, the definition of $t$ could be confusion. Let $\tau$ denote a time range in a week. Then, $\bm{D}_{w,\tau}$ denote the population density data set in week $w$ and time range $\tau$. Let $\tau^1, \tau^2$ be the latest periods' time ranges of two strategies and $\tau^3$ be the time range we need to predict. Formula~\ref{func:lstm} is improved as:
\begin{equation}
\label{func:dual}
F_{fc}(F_{LSTM}^1(\bm{D}_{w,\tau^1}),F_{LSTM}^2(\bm{D}_{w,\tau^2}),\bm{D}_{w-1,\tau^3})=\bm{D}_{w,\tau^3}
\end{equation}

\section{Parallel Architecture}\label{sec:parall}

\begin{figure}[!htbp]
\centering
\includegraphics[width=0.9\linewidth, clip=true, trim=10 0 0 0]{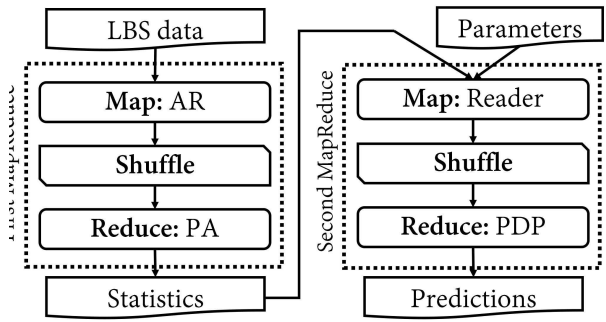}
\caption{The parallel architecture contains two MapReduce jobs. First MapReduce uses \emph{address resolution module} as map and \emph{population analysis module} as reduce. After the shuffle, the logs are grouped by the hospital they belong to and sorted by their time tags. Second MapReduce uses the shuffle to pair the model parameter and the statistics for the \emph{population prediction module}.}
\label{fig:parallel-structure}
\end{figure}

To obtain the (nearly) real-time population density of all hospitals in Beijing, our application has to process 182GB data each day. Considering the number of active mobile users in the daytime is far more than that in the nighttime, the application demands the capacity of processing 20GB data in one hour. To this end, we designed a parallel architecture based on \emph{Hadoop}. The architecture is composed of two \emph{MapReduce} jobs, as shown in Figure~\ref{fig:parallel-structure}.

Since population analysis of each hospital is independent, it can be implemented in parallel. We utilize AR module as \emph{Map} and PA module as \emph{Reduce}. The only condition is that all data of one hospital have to be processed by the same slave node in the \emph{MapReduce} job since an individual's behavior is characterized by all of his logs that relate with the hospital. In addition, crowd counting also requires all data of the hospital. However, data obtained from Baidu LBS log database is sorted by time tag. To resolve this contradiction, we customize \emph{Hadoop}'s \emph{Partition} and \emph{Comparison} functions in the \emph{Shuffle} phase and ensure all the data belonging to the same hospital will be distributed to the same \emph{Reducer} (i.e. slave node) and sorted by the time tag.

The second \emph{MapReduce} job is for PDP module. This is also an independent work for different hospitals, so we utilize it as \emph{Reduce}. As PDP module for different hospital entails different parameters, slave nodes have to load corresponding parameters and history data. By using the similar strategy, \emph{Map} job simply reads and exports all parameters and history data. Meanwhile, \emph{Hadoop}'s \emph{Partition} and \emph{Comparison} functions in the \emph{Shuffle} phase will deliver these data correctly to each \emph{Reducer}.

The first \emph{MapReduce} job runs once per hour and the second \emph{MapReduce} job runs once every two hours since in Section~\ref{sec:PDP} the smallest step size is $2$. Note that the \emph{counting lists} and \emph{blacklists} are preserved to disk each time as PA job is finished and will be reloaded by next hour's job. So, the PA module can be seen as working continuously.

\section{Implementation}\label{sec:implementation}

In this section, the implementation detail of our application will be introduced.

\begin{figure*}[!htbp]
\centering
\subfigure[]{
  \includegraphics[width=0.3\linewidth, clip=true, trim=0 0 0 0]{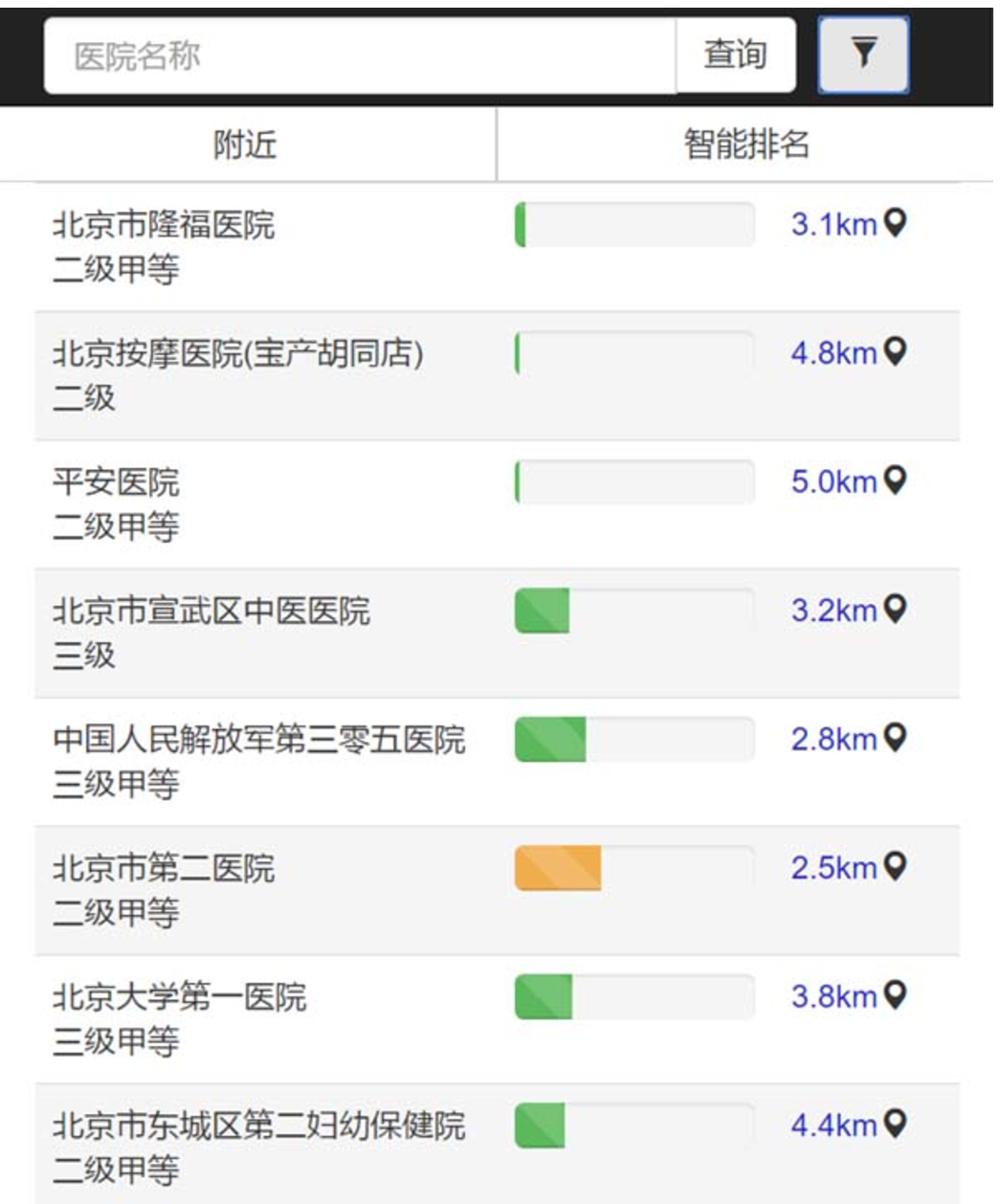}
  \label{fig:app1-1}
}
\subfigure[]{
  \includegraphics[width=0.3\linewidth, clip=true, trim=0 0 0 0]{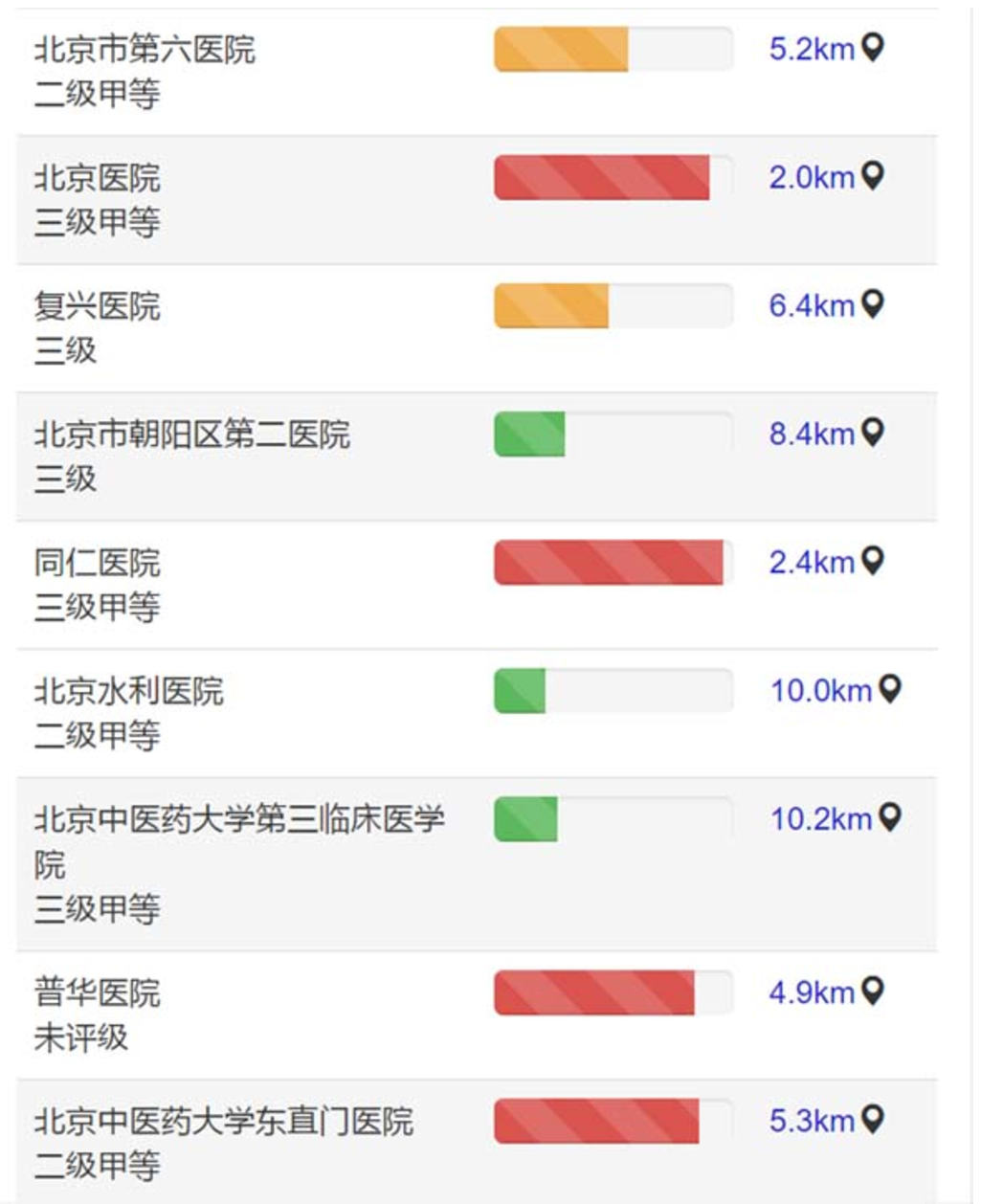}
  \label{fig:app1-2}
}
\subfigure[]{
  \includegraphics[width=0.3\linewidth, clip=true, trim=0 0 0 0]{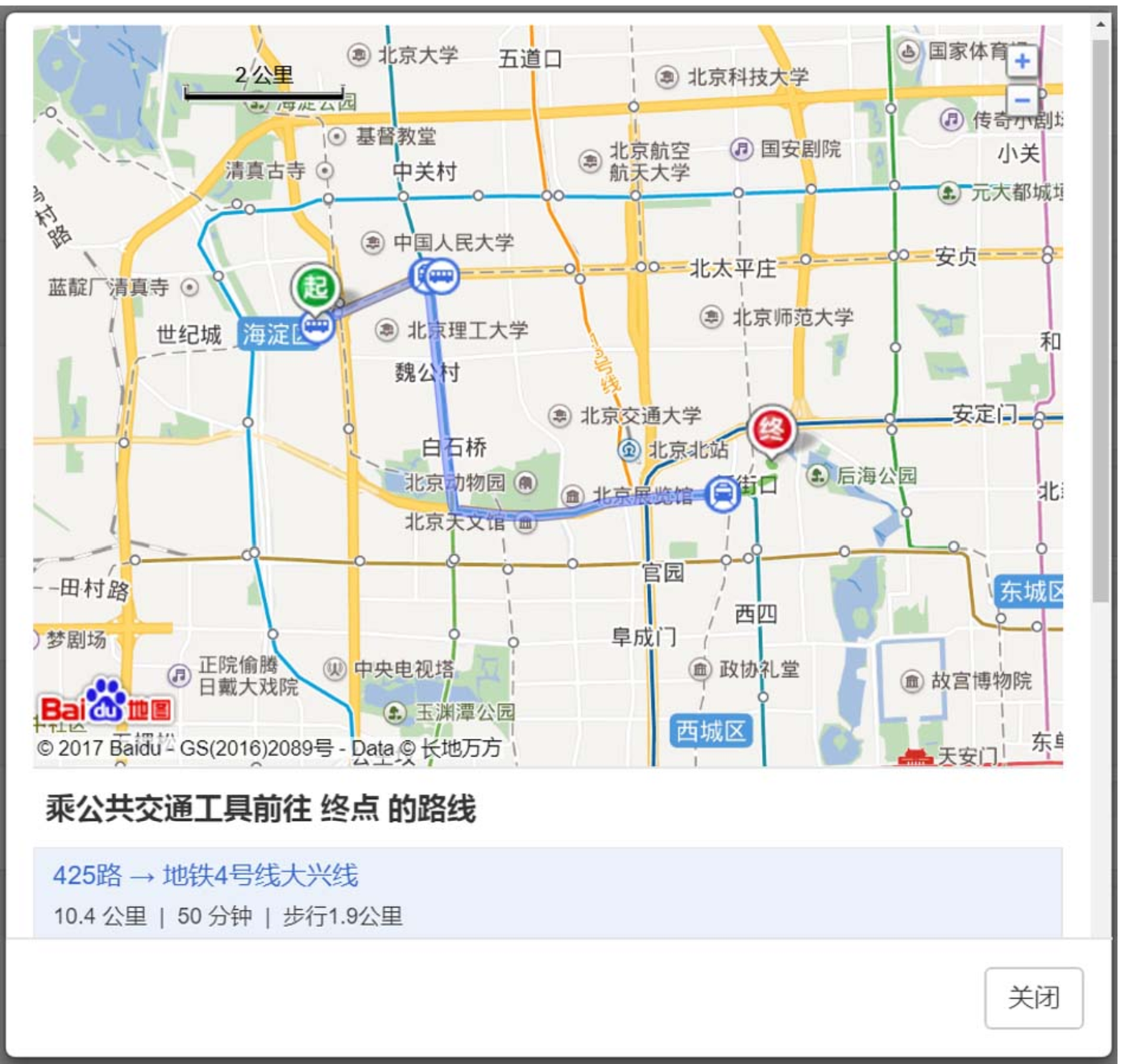}
  \label{fig:app3-1}
}
\subfigure[]{
  \includegraphics[width=0.3\linewidth, clip=true, trim=0 0 0 0]{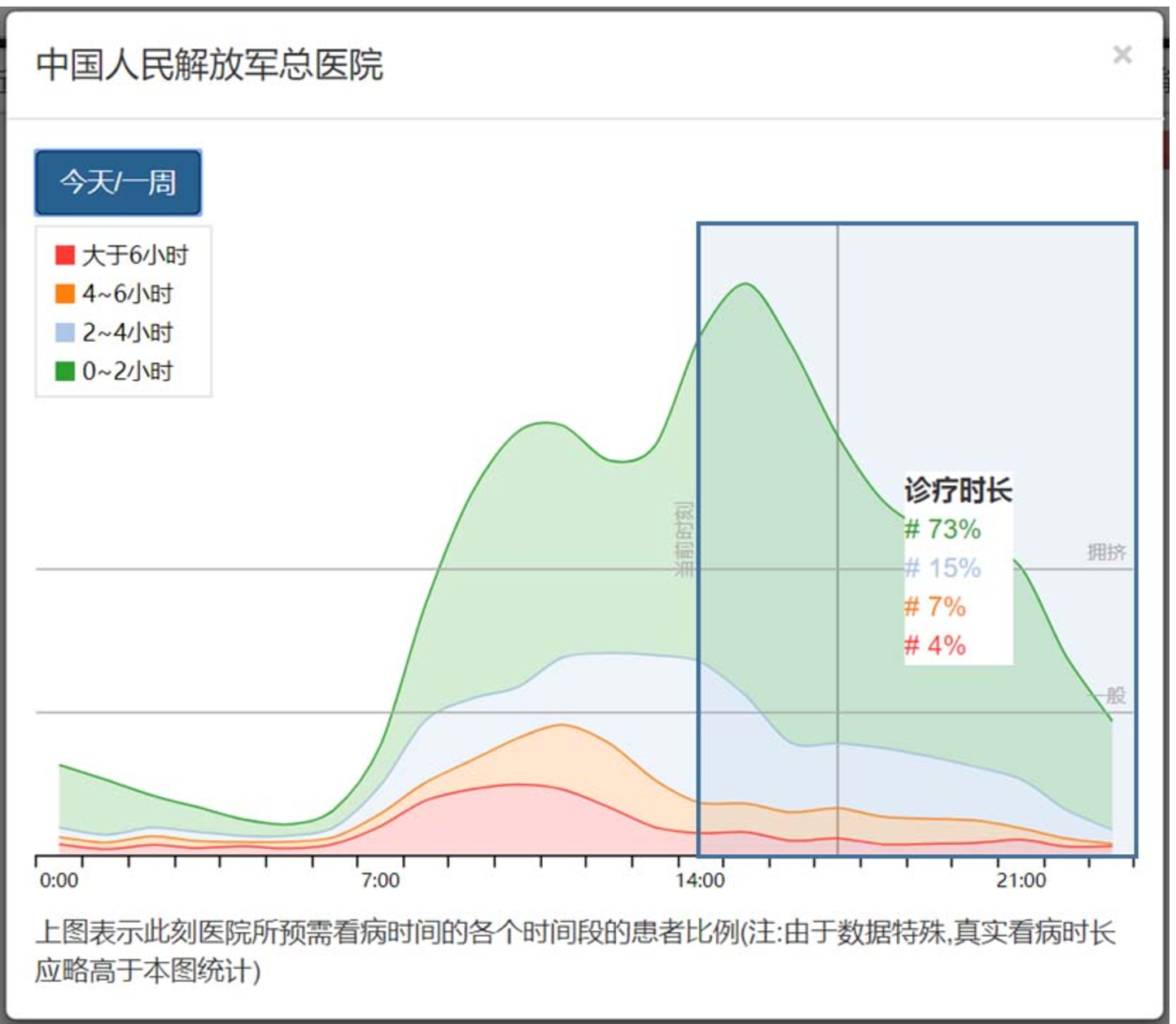}
  \label{fig:app2-1}
}
\subfigure[]{
  \includegraphics[width=0.3\linewidth, clip=true, trim=0 0 0 0]{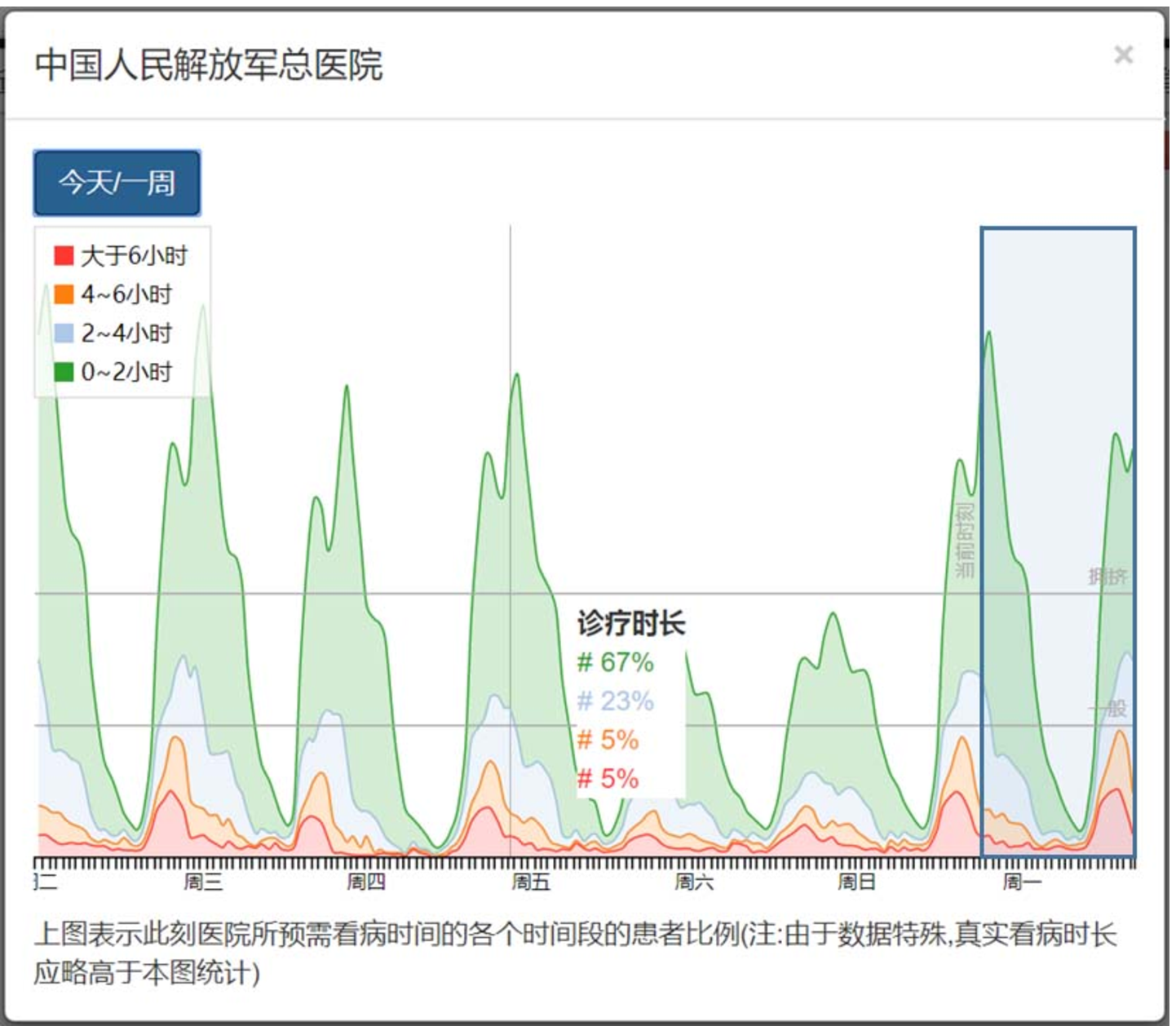}
  \label{fig:app2-2}
}
\subfigure[]{
  \includegraphics[width=0.3\linewidth, clip=true, trim=0 0 0 0]{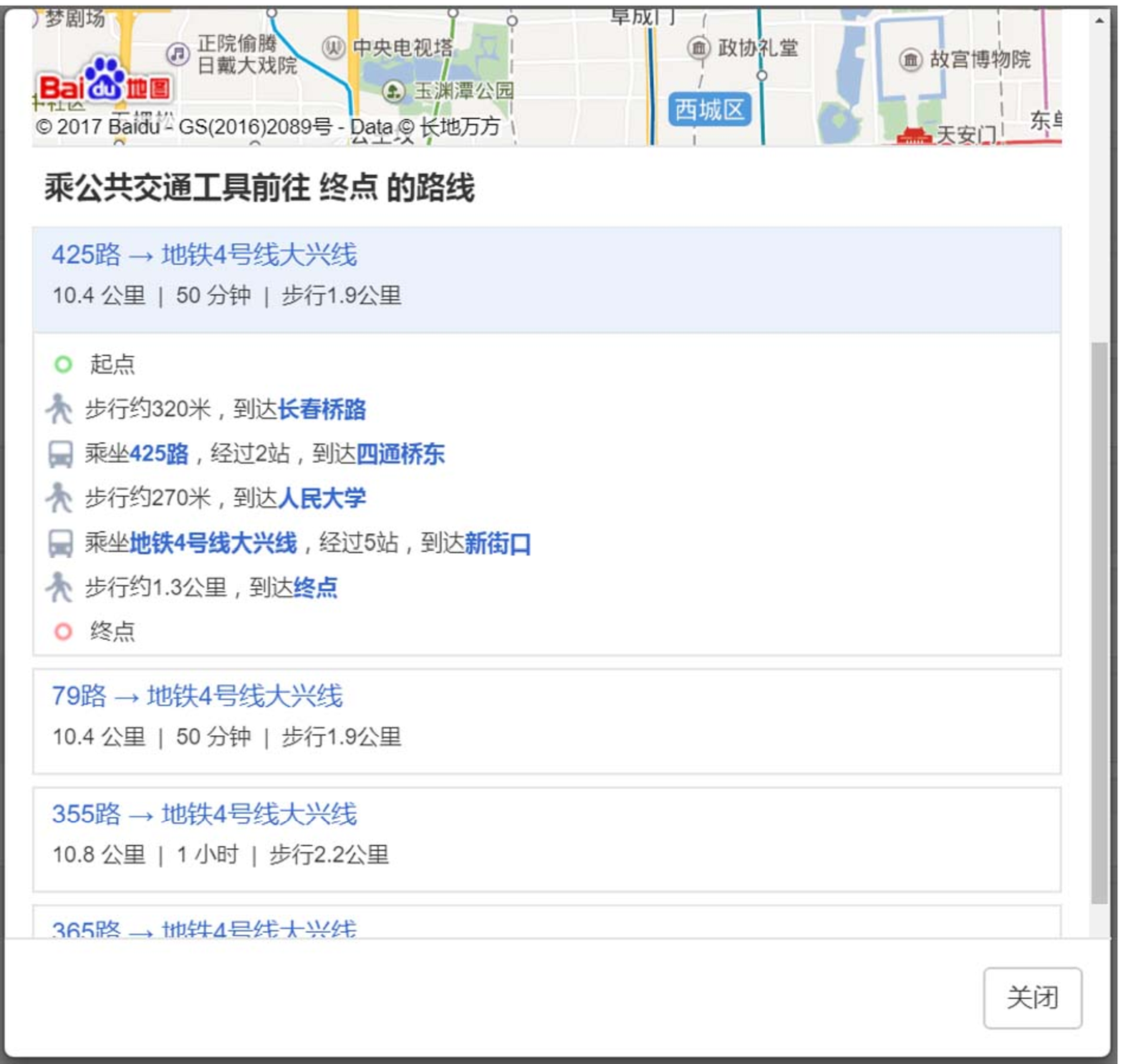}
  \label{fig:app3-2}
}
\caption{Screenshots of our application. Since current major users of the application are Chinese, the application is in Chinese. Fig. \ref{fig:app1-1} and Fig. \ref{fig:app1-2} show the recommendation list. Fig. \ref{fig:app2-1} and Fig. \ref{fig:app2-2} show the daily and weekly density-time curves. Fig. \ref{fig:app3-1} and Fig. \ref{fig:app3-2} show the path plan function.}
\label{fig:app}
\end{figure*}

\subsection{Dataset}\label{subsec:im-DP}
In this study, we used three crucial datasets, i.e., Baidu LBS request logs, Baidu points of interests (POI) and hospitals' basic information. 

\textbf{Baidu LBS request logs.} The attributes of Baidu LBS request logs cover \emph{anonymous user ID} (UID), \emph{latitude}, \emph{longitude}, \emph{positioning accuracy}, and \emph{time tag}. The quantity of LBS logs in Beijing every day is in the order of $10^8$. Although this dataset cannot represent all the demography in each hospital, such as very young and very old people, or those who do not use smartphones, massive data like this is sufficient to approximate the population density of hospitals, which is the focus of our study.

\textbf{Baidu POI.} This dataset includes POI name and boundary coordinates. We gather the geographic information like boundary and area of hospitals from this dataset.

\textbf{Hospitals' basic information} This dataset contains hospitals' name, official class and the number of doctors. We established this dataset by ourselves through merging the information grabbed from three reliable websites\footnote{http://www.xywy.com/, http://www.haodf.com/, and http://www.39.net/}.

\subsection{Model}\label{subsec:im-model}
In Section~\ref{subsec:AR} we mentioned that for a fast query, the map between grids' IDs and their semantic information is stored. In the application, we use a hash table to store the map and we only stored the grids located in a hospital to save space. In this case, as the grids not located in any hospital is far more than those located in one, there might be serious hash collisions and make AR module misclassify the logs not located in any hospital. An addition hash function is applied to generate a fingerprint to compensate the flaw. Though only hospital grids are stored, the amount of them is still enormous. To meet the balance between checking and space efficiency, we utilize Cuckoo Filter~\cite{fan2014cuckoo} which is an amazing hash structure with $O(1)$ checking complexity and more than $95\%$ memory efficiency. 

To obtain a more reliable prediction, we improved $\bm{D}_{w-1,\tau^3}$ in Formula~\ref{func:dual}. It is replaced by $\bm{\overline{D}}_{\tau^3}$ averaged in three months.

\subsection{Application}

In the application, instead of frankly showing the crowd counting number, we mix the population density data and the hospital basic information such as hospital area and doctor number to provide a more practical and user-friendly population density level as the colorful bars in Figure~\ref{fig:app1-1} and Figure~\ref{fig:app1-2} show. The mixing strategy is empirically formulated.

Figure~\ref{fig:app} shows screenshots of our application. The recommendation list sorted by the merged information includes hospitals' name and level, outpatient densities level and the distance between the user and the hospital, as Fig. \ref{fig:app1-1} and Fig. \ref{fig:app1-2} show. Here, the merging strategy is also empirical. In addition, we also provide daily and weekly density-time curves. On the curve, users can check detail ratios of outpatients with different resident time at any point. This function showed in Fig. \ref{fig:app2-1} and Fig. \ref{fig:app2-2}. The curve in the blue frame denotes the future trend predicted by PDP module. From these two figures it can be seen that although the instability of the LBS data makes the curve of long-duration patients noisy, they still provide some suggested guidance. Moreover, by clicking the distance in the rank list, users can also check the path plan and the public transit route to the hospital, showed in Fig. \ref{fig:app3-1} and Fig. \ref{fig:app3-2}.

\section{Experiment}

\begin{figure}[!htbp]
\centering
\subfigure[]{
  \includegraphics[width=0.45\linewidth, clip=true, trim=0 0 0 0]{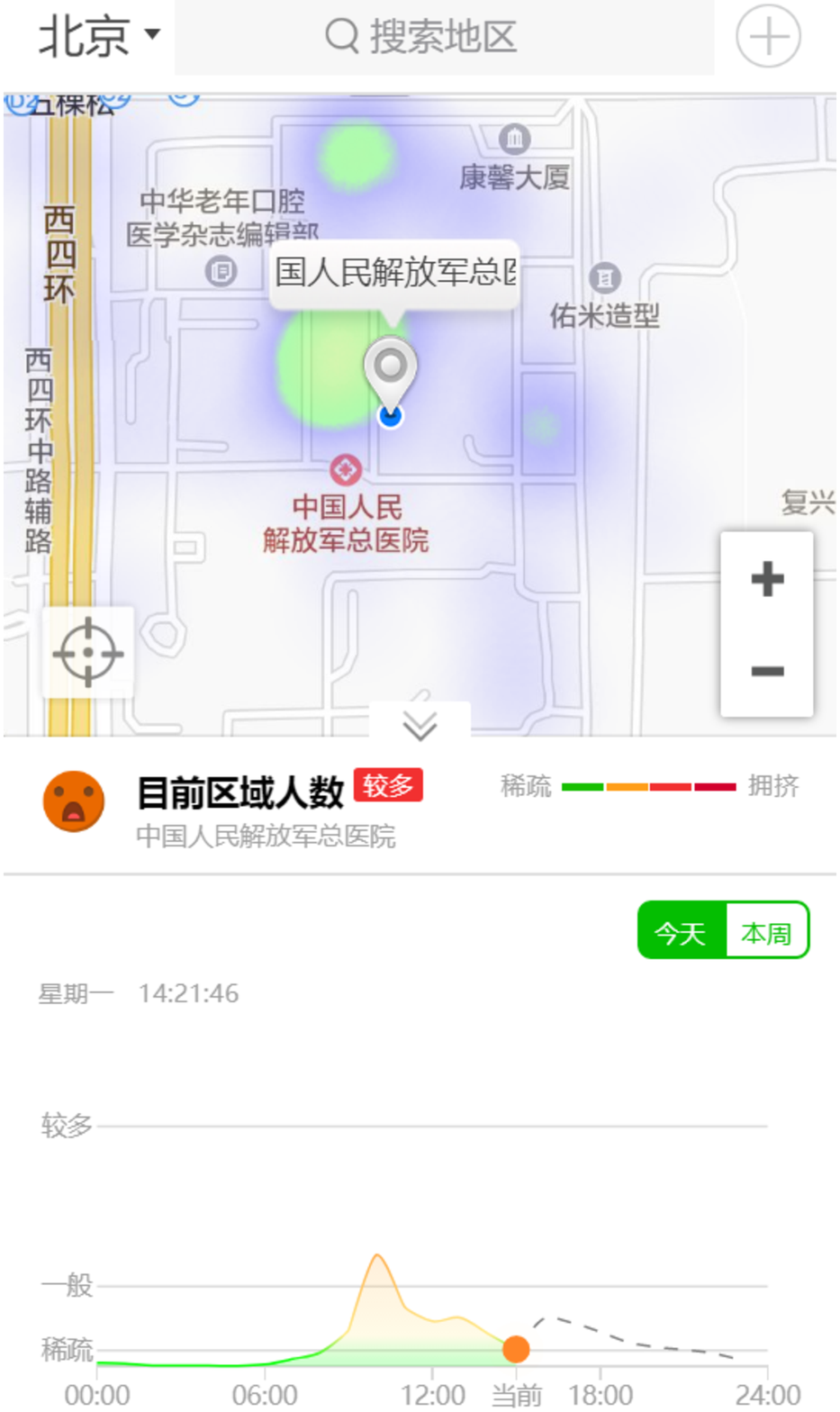}
  \label{fig:wechat-1}
}
\subfigure[]{
  \includegraphics[width=0.45\linewidth, clip=true, trim=0 0 0 0]{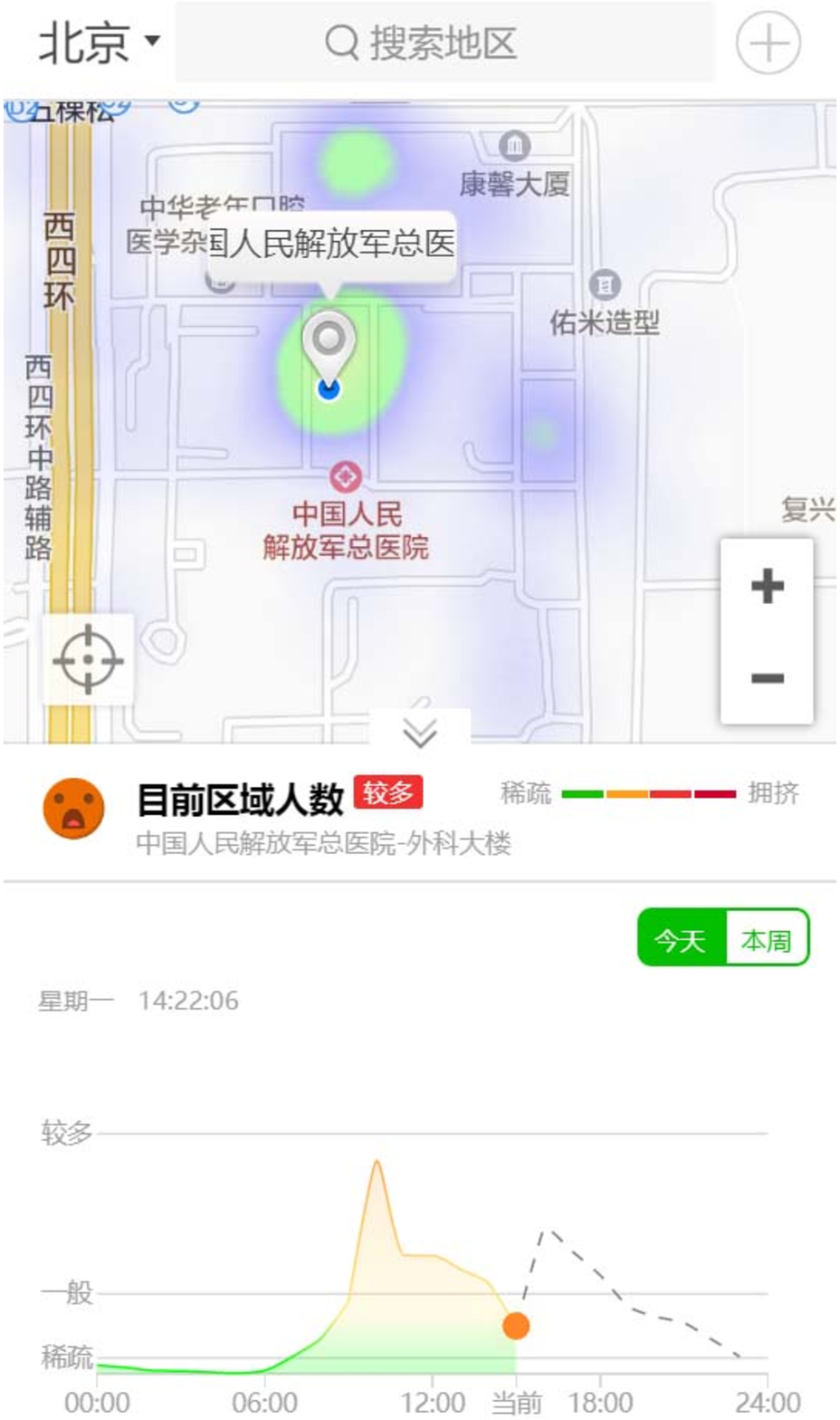}
  \label{fig:wechat-2}
}
\caption{The screenshots of Wechat City Heat Map.}
\label{fig:wechat}
\end{figure}

In this section, we evaluate the quality of the estimation and prediction of population density separately. Since the true number of outpatients in every hospital in Beijing is hard to count, we qualitatively evaluate the performance of our population analysis approach by comparing the density-time curve with Wechat City Heat Map. For the accuracy of population density prediction, as the input of PDP module is the statistics estimated by PA module, we use that as ground truth.

\subsection{Population Density Analysis}

Figure~\ref{fig:wechat} shows the density-time curves provided by Wechat City Heat Map. The result of our method on the same hospital at the same time can be seen in Figure~\ref{fig:app2-1}. Wechat City Heat Map can only show the population density of one point and the results of two close points in the same hospital are quite different, while our result is naturally generated on the whole area of the hospital. Furthermore, our result provides a finer-granular density of outpatients instead of the entire population. Therefore, our approach can present a more instructive information in practice.

\subsection{Population Density Prediction}

Since our aim is to recommend proper hospitals to potential patients, we also concern the rank of predicted values. So we estimate the quality from two perspectives: \emph{Spearman Rank Correlation Coefficient} and \emph{Relative Error} of the prediction. The \emph{Spearman Correlation Coefficient} is defined as the Pearson correlation coefficient between the ranked variables~\cite{myers2010research}. The definition of \emph{Relative Error} is:
\begin{equation}
\delta_{y,g}=\frac{|y-g|}{g}
\end{equation}
where $y$ is the prediction and $g$ is the statistical ground truth. Table \ref{tab:ex} shows the results.
\begin{table}[htbp]
\centering
\caption{SRCC indicates \emph{Spearman Rank Correlation Coefficient} and RE indicates \emph{Relative Error}. The results are averaged across 460 times predictions over 113 hospitals. Higher SRCC and lower RE are preferred as indicated by the arrows in the table.}
\begin{tabular}{lcc}
    \toprule
    &mean&variance\\
    \midrule
    SRCC&0.857$^\uparrow$&0.071\\
    RE&0.224$_\downarrow$&0.098\\
    \bottomrule
\end{tabular}
\label{tab:ex}
\end{table}

Considering one prediction includes several hours' population density trend, we also evaluate the internal stability. In Formula~\ref{func:dual}, consider that $\tau^3$ includes $n^h$ hours. Let $h$ be the index of an hour in the period $\tau^3$, $h\in\{1,2,..,n^h\}$, and $y_h$ be the predicted population density of hour $h$. A prediction is denoted by $\bm{Y}=\{y_1,y_2,..,y_{n^h}\}$. The corresponding ground truth is $\bm{G}={g_1,g_2,..,g_{n^h}}$. Let the current time is $\dot{t}$, the time of $y_h$ is $\dot{t}'$. Normally, $\tau^3$ starts from $\dot{t}+1$. Thus, it is obvious that $h=\dot{t}'-\dot{t}$. As $h$ increases, we evaluate if $\delta_h$ will increase, in other words, if the precision decline. Here $\delta_h$ is the abbreviation of $\delta_{y_h,g_h}$.

\begin{figure}[htbp]
\centering
\includegraphics[width=1\linewidth, clip=true, trim=130 50 115 50]{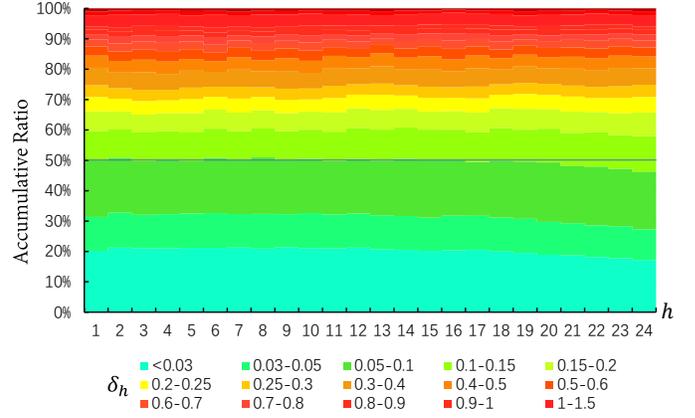}
\caption{The results of the experiment on PDP module. The $x$-axis represents $h$. Relative error $\delta_h$ is sliced into 17 ranges represented by 17 different colors from cyan to red.}
\label{fig:lstm-ex}
\end{figure}

In the experiment, we set $n^h=24$. Figure~\ref{fig:lstm-ex} shows the result. For better illustration, we use different colors to represent different ranges of $\delta_h$. And the area of each color indicates the ratio of results within corresponding precision range. From the figure, it can be seen that about $50\%$ data has relative error $\delta \le 0.1$. As $h$ increases, further, our model only has a subtle decrease in performance.

\section{Case Study}\label{sec:cs}

\begin{figure}[htbp]
\centering
\subfigure[]{
	\label{fig:density-time-d}
	\includegraphics[width=1\linewidth, clip=true, trim=50 70 50 60]{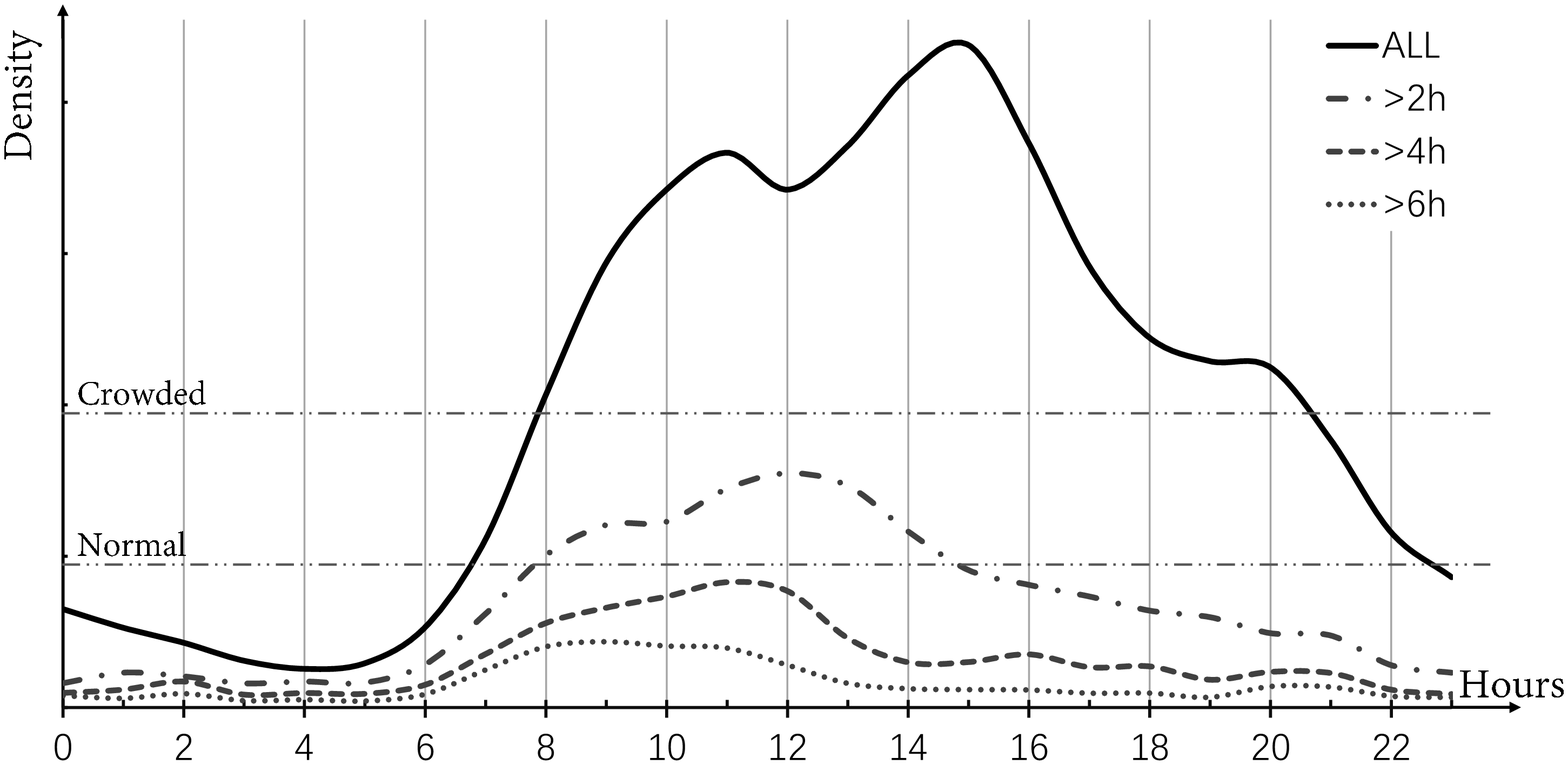}
	}
\subfigure[]{
	\includegraphics[width=1\linewidth, clip=true, trim=30 60 30 55]{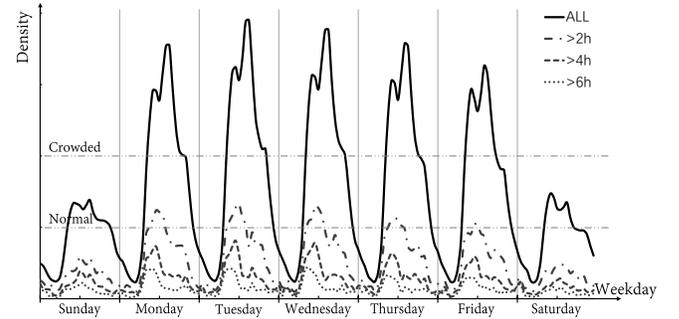}
	\label{fig:density-time-w}
	}
\caption{The density-time curves. The densities are averaged across 3 months. Thresholds of 'Normal' and 'Crowded' are both empiric values. Four curves in each figure respectively represent aggregated density of outpatients staying over 2 hours, over 4 hours, and over 6 hours.}
\label{fig:density-time}
\end{figure}

To find out the varying patterns of population density, we analyze two representative TOP-CLASS hospitals in Beijing. Figure~\ref{fig:density-time-d} illustrates the Tuesday's density-time curve of outpatient at Peking University Third Hospital, and Figure~\ref{fig:density-time-w} is the weekly density-time curve of outpatient at Peking University People's Hospital. 

As shown in Figure \ref{fig:density-time}, population density has an obvious daily cycle with a clear valley between 11 a.m. to 1 p.m. We speculate that it is related to the hospitals' worktime. Most of the hospitals have a noon break from 12 a.m. to 1:30 p.m. Generally, hospitals stop taking appointments for forenoon at 11:30 a.m., which means the people waiting for treatment in the morning begin to decrease at around 11 a.m. Then people in the hospital start leaving for lunch. After worktime has finished, population density touches bottom. Starting from 1 p.m., patients return to the hospital waiting for doctors or registration, leading to the gradual increase of population density.  Another drop of population density starts from 3 pm, as shown in Figure \ref{fig:density-time-d}. The reasons are that {\bf{a)}} most of the hospitals stop taking appointments at half past 4 p.m. Since the registration does not have noon break and the number of appointments for one day is limited, appointments for many hot departments are sold out at about 3 or 4 p.m. {\bf{b)}} Hospital outpatient clinics are usually closed at half past 5 p.m. The patients may not have enough time for some medical examinations if arriving after 3 o'clock. 

In the long term, it can see from Figure \ref{fig:density-time-w} that the number of patients decreases from Tuesday to Friday. Despite this pattern is not strictly followed by all the hospitals, every Tuesday and Friday are still the peak and valley of the crowd for the majority of the hospitals. Several potential underlying causes are that {\bf{a)}} many medical examinations are not provided over the weekends. Thus, these examinations' appointments made at weekend or Friday are often postponed to Monday or Tuesday. {\bf{b)}} Meanwhile, there are many outlanders seeking outstanding doctors in Beijing. They often prefer to go to the hospital at the first half of the week because they might have to make appointments for some medical examinations. Making it early is more likely to get a comprehensive examination by utilizing this whole week, which can save the cost of booking room. A direct impression from these two factors is that hospitals should be the most congestion on Monday. Therefore,  many people steer clear of going to hospital on Monday, leading to that Tuesday becomes the most crammed day. Overall, for the TOP-CLASS hospitals, the population densities are more reasonable in the latter part of the week.

\section{Conclusion}\label{sec:conclu}
We presented a novel approach for hospital population density estimation based on LBS big data. With the extraction of the LBS request logs around the hospitals, it can effectively estimate the current population density and predict the trend in each hospital based on big data analytics. We expect our application of the approach can guide outpatients to choose appropriate hospitals based on different criteria. Since our approach circumvents the data collection issue, it can be freely applied on a large spatial scale such as a city, a province, even the entire country.

The proposed approach can directly benefit from more advanced recommendation strategies and more accurate and stable LBS data. In addition, we use Hadoop for parallel computing due to the limitation of the experiment environment. Better speed-up performance might be achieved by using Spark in the future. More broadly, the use of empirical thresholds and rules brings limitations of the approach. It is worthwhile to develop more flexible strategies to expand the capacity of the approach and explore its potential of balancing the human distributions in many other similar scenarios such as resorts, gyms, supermarkets and other public areas. And it is also imperative to develop an effective evaluation approach for such extensive crowd counting problem.

\bibliographystyle{ieeetr}
\bibliography{ref}

\end{document}